\documentclass[english]{article}
\usepackage[superscript]{cite}
\usepackage[utf8]{inputenc}
\usepackage[T1]{fontenc}
\usepackage{hyperref}
\usepackage{babel}
\usepackage{color}
\usepackage{amsmath}
\usepackage{xspace}
\usepackage{graphicx}
\usepackage{fancyhdr}
\usepackage{url}
\usepackage{multirow}

\makeatletter \renewcommand\@biblabel[1]{#1.} \makeatother

\makeatletter
\newcommand*{\etc}{%
    \@ifnextchar{.}%
        {etc}%
        {etc.\@\xspace}%
}
\newcommand*{\etal}{%
    \@ifnextchar{.}%
        {et al}%
        {et al.\@\xspace}%
}
\makeatother

\newcommand{\revision}[1]{#1}

\begin{document}

\title{Dataset of segmented nuclei in Hematoxylin and Eosin stained histopathology images of 10 cancer types}

\author{Le Hou\textsuperscript{2}, Rajarsi Gupta\textsuperscript{1}, John S. Van Arnam\textsuperscript{1}, Yuwei Zhang\textsuperscript{1}, \\
Kaustubh Sivalenka\textsuperscript{2}, Dimitris Samaras\textsuperscript{2}, Tahsin M. Kurc\textsuperscript{1}, \\
Joel H. Saltz\textsuperscript{1,{*}}}

\date{}

\maketitle
\thispagestyle{fancy}
\begin{center}
1. Biomedical Informatics Department, HSC L3-045\\
Stony Brook Medicine, Stony Brook University \\
Stony Brook, NY 11794. USA\\
\vspace{0.1cm}
2. Computer Science Department, 203C New Computer Science Building\\
Stony Brook University\\
Stony Brook, NY 11794. USA\\
\vspace{0.1cm}
{*}corresponding author: Joel Saltz (joel.saltz@stonybrookmedicine.edu)
\end{center}

\begin{abstract}
The distribution and appearance of nuclei are essential markers for the diagnosis and study of cancer. Despite the importance of nuclear morphology, there is a lack of large scale, accurate, publicly accessible nucleus segmentation data. To address this, we developed an analysis pipeline that segments nuclei in whole slide tissue images from multiple cancer types with a quality control process. We have generated nucleus segmentation results in 5,060 Whole Slide Tissue images from 10 cancer types in The Cancer Genome Atlas. One key component of our work is that we carried out a multi-level quality control process (WSI-level and image patch-level), to evaluate the quality of our segmentation results. The image patch-level quality control used manual segmentation ground truth data from 1,356 sampled image patches. The datasets we publish in this work consist of roughly 5 billion quality controlled nuclei from more than 5,060 TCGA WSIs from 10 different TCGA cancer types and 1,356 manually segmented TCGA image patches from the same 10 cancer types plus additional 4 cancer types. Data is available at \url{https://doi.org/10.7937/tcia.2019.4a4dkp9u}
\end{abstract}

\section*{Background \& Summary}

Digital pathology images are obtained via a series of processes: tissue slicing, staining, image capturing and digitization. The resolution of these images is usually at multi-gigapixel level. A single tissue slide typically contains around a million nuclei. The appearance, shape, texture, and morphological features of nuclei depend on the tissue type excised from an organ, cancer type, cell type, and many other factors. The comprehensive detection, segmentation, and classification of nuclei are core analysis steps in many histopathology image analysis tasks \cite{gurcan2017digital,colen2014nci,xie2015beyond,cooper2012digital,saltz2017towards,bayramoglu2016transfer,xu2016stacked,wang2016subtype,chen2017dcan,zhang2017deep,yang2017suggestive,bai2017deep,kumar2017dataset,murthy2017center,hou2019sparse,naylor2018segmentation}. Segmentation of nuclei is the first step in extracting interpretable features that provide valuable diagnostic and prognostic cancer indicators~\cite{cooper2010integrative,cooper2012integrated,parmar2015radiomic,gillies2015radiomics,aerts2014decoding}, and thus is a crucial step for precision medicine \cite{national2011toward,collins2015new}.  The Cancer Genome Atlas (TCGA) program was a decade long, large scale National Cancer Institute led research effort that molecularly characterized over 20,000 primary cancer and matched control samples spanning 33 cancer types. Diagnostic whole slide images were captured for a large fraction of TCGA patients. Deidentified whole slide images, linked to molecular and clinical information are frequently accessed and analyzed publicly available information.  TCGA whole slide Pathology images have been employed in many Cancer research efforts as well as in many digital Pathology methodology studies; 
Cooper et al.~\cite{cooper2018pancancer}, for instance, describes examples of how TCGA whole slide images were used in integrative TCGA studies.

Current efforts to generate publicly accessible nuclear segmentation datasets in Hematoxylin and Eosin (H\&E) stained whole slide images have been 
at much smaller scales than our work. Kumar et al. \cite{kumar2017dataset} collected a dataset of nucleus segmentation in seven cancer disease sites. This dataset is used as the MICCAI 2018 MoNuSeg challenge \cite{kumar2017multi} in which the training set contains 30 image patches containing around 22,000 nuclear boundary annotations. The MICCAI 2015 to MICCAI 2018 Segmentation of Nuclei challenge \cite{vu2019methods} training sets contain around 6,000 nuclear boundary annotations. \revision{The extended PanNuke dataset \cite{gamper2020pannuke,gamper2019pannuke} (currently the largest available dataset) contains 205,343 semi-automatically segmented nuclei in 481 patches sampled from 19 tissue types.} Other datasets \cite{janowczyk2016deep,wienert2012detection,irshad2014crowdsourcing,Drelie08-298} have similar or smaller numbers of segmented nuclei. For these existing datasets, training patches are usually stain-balanced, well digitized, and do not contain rare textures. However, in real world applications, the appearance of nuclei can be affected by a number of staining and imaging conditions: extremely high cellularity and nuclear pleomorphism, slightly out-of-focus, folding tissue, imbalanced H\&E staining, \etc. Existing experiments \cite{hou2019robust} showed that Convolutional Neural Networks (CNNs) generalize sub-optimally in unseen cancer types (cancer types that do not have training data). Therefore, training segmentation CNNs on existing datasets naively yields poor segmentation results in WSIs \cite{hou2019robust}.

We aimed to accurately segment nuclei in WSIs of multiple cancer types. For this purpose, we leveraged a state-of-the-art nucleus segmentation Convolutional Neural Network (CNN) that our group recently reported \cite{hou2019robust}.  Our approach has two advantages: (1). It generalizes well in cancer types that do not have training data: it improves the robustness of the segmentation network by synthesizing training data of every cancer type (2) The method is computationally efficient - this was critical given our goal of computing segmentation results for over 5,000 WSIs. Given our ability to produce large scale synthetic training data, a small U-net CNN \cite{ronneberger2015u} was able to generate accurate instance-level segmentation results in around \revision{\textbf{3 GPU hours per WSI}}. Computationally expensive networks such as the Mask R-CNN \cite{he2017mask} would achieve similar or worse across-cancer type generalization performance but in over 30 GPU hours per WSI. By combining three real training datasets \cite{kumar2017dataset,vu2019methods} and a large scale synthetic dataset of 500,000 image patches, we train a U-net that has two output heads: one for nuclear center detection and one for nuclear material segmentation. We finally applied the watershed method \cite{beucher1994watershed,bai2017deep} on detected centers and segmentation results, to output instance-level segmentation.

No existing automatic segmentation models give perfect results. Visually assessing segmentation results over 5,000 WSIs would take more than 200 human hours (more than 2.5 minutes per WSI) which is very time consuming. Instead, we apply the following methods \textbf{for quality control and data validation:}
\begin{description}
    \item[Patch-level quantitative evaluation] We manually segmented nuclei in 1,356 patches and leveraged this to quantitatively evaluate our 5,000+ WSI segmentation dataset. In particular, we measured the segmentation overlap using Dice scores, and the instance-level segmentation/detection quality using Instance-Dice scores \cite{vu2019methods} and the nuclei count correlation scores.
    \item[Random segmentation region checking and WSI-level quality control] (1) We sampled 15 patches per WSI, and visually assessed and manually marked patches with what we considered to be adequate segmentation results (both precision and recall are at least 75\%). (2) We identified WSIs that have unusual segmentation statistics (too few/many segmented nuclei \etc), then visually assess segmentation data in them, and marked slides that have unacceptable segmentation (less than 80\% of the slide both precision and recall are at least 75\%). In these ways, we categorized WSIs into groups with different segmentation quality levels. 
    
\end{description}

\begin{figure}
\begin{center}
   \includegraphics[width=1.0\linewidth]{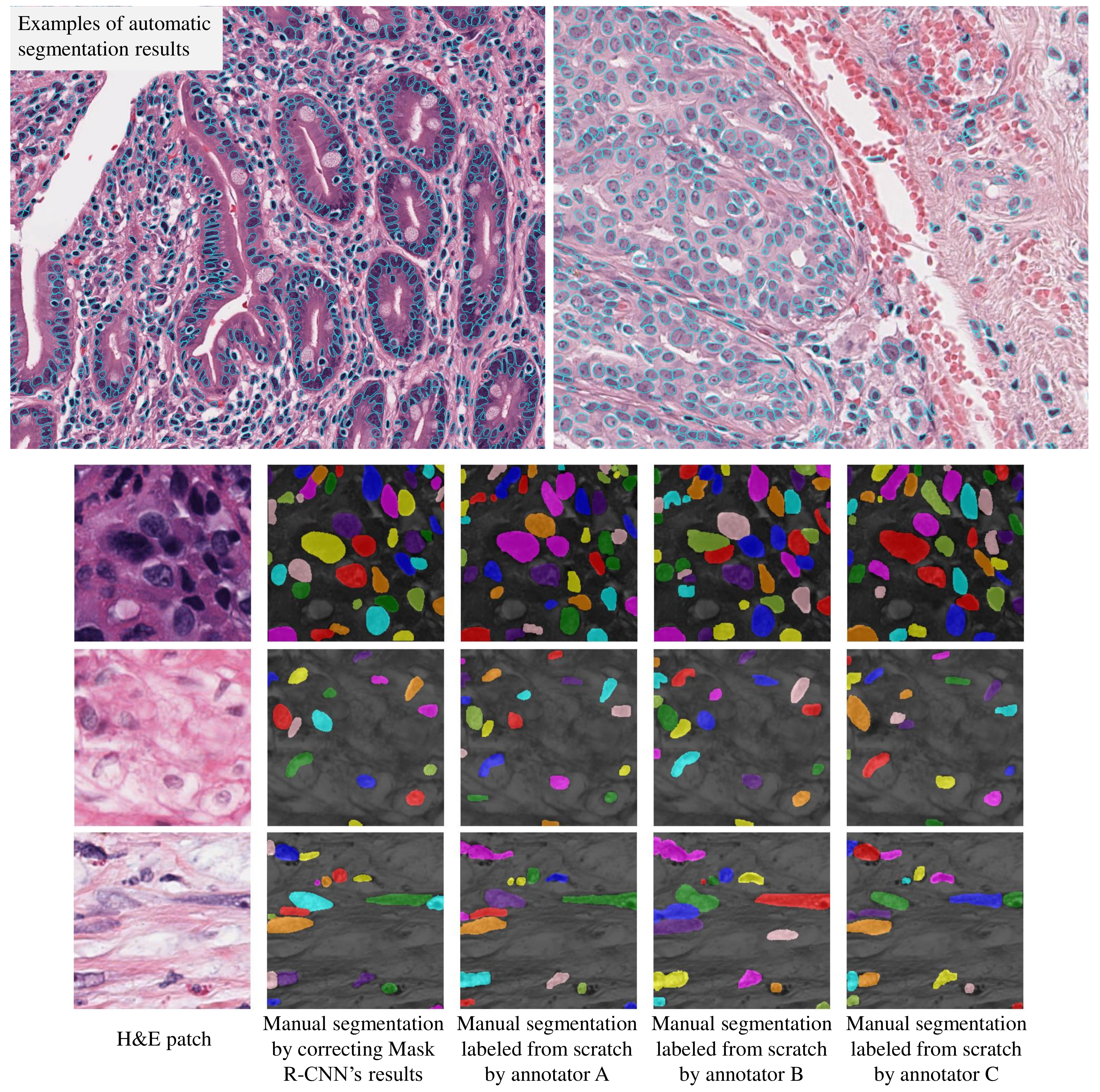}
\end{center}
   \caption{Samples of our data. (1). Automatic segmentation results on 5,060 WSIs (samples in top row), summarized in Table \ref{tab:data_results}. (2). Manual segmentation data on over 1,356 patches (samples in bottom rows). Coloring of nuclear masks is for visualization only: it differentiates individual nuclei. We collect a large number of patches with labels for validating the segmentation results.}
\label{fig:wsi-seg-results}
\end{figure}

Using the patch-level manual segmentation data in 14 different TCGA cancer types, we quantitatively evaluated segmentation data. We judged 10 of the 14 cancer types to have nuclear segmentation result quality worthy of publication and data release.  We thus release the following validated data as \textbf{our contributions}:
\begin{enumerate}
    \item The automatic nucleus segmentation dataset contains 5,060 segmented slides in 10 TCGA cancer types, summarized in Table \ref{tab:data_results}. This represents approximately 5 billion segmented objects. This large scale segmentation data for TCGA slides is very important, since characteristics of nuclei are essential for the diagnosis and study of cancer.
    \begin{enumerate}
        \item We apply per-WSI level quality control and categorize WSIs into groups with different segmentation quality levels. We identified 576 slides with suboptimal segmentation results. We filter out those WSIs for further analysis (although we still release the data for completeness).
        \item Based on our patch-level quantitative assessment, compared to manual segmentation, in every cancer type, \revision{the nucleus segmentation data has an average Dice coefficient of least 77\%, and an average instance level Dice coefficient \cite{vu2019methods} of at least 62\%. These results are similar to the inter-annotator agreement in our experiments.}
    \end{enumerate}
    \item Manual segmentation labels on 1,356 patches of 256 $\times$ 256 pixels (64 $\times$ 64 $\mu m^{2}$) uniformly distributed in 14 cancer types. Two pathologists collaborated with three graduate students employed results from Mask R-CNN as a base to generate segmentation labels.
\end{enumerate}
Examples of both datasets are shown in Figure \ref{fig:wsi-seg-results}.

\begin{table}
  \centering
  \begin{tabular}{l p{5.5cm} r r}
    \hline
    & & \#. slides & \#. slides \\
    Abbre. & Cancer type & in total & failed QC \\
    \hline
BLCA & Urothelial carcinoma of the bladder & 380 & 14\\
BRCA & Invasive carcinoma of the breast & 1,096 & 88 \\
CESC & Cervical squamous cell carcinoma and endocervical adenocarcinoma & 249 & 54 \\
GBM & Glioblastoma Multiforme & 772 & 40 \\
LUAD & Lung adenocarcinoma & 540 & 59 \\
LUSC & Lung squamous cell carcinoma & 431 & 35 \\
PAAD & Pancreatic adenocarcinoma & 190 & 11 \\
PRAD & Prostate adenocarcinoma & 387 & 19 \\
SKCM & Skin Cutaneous Melanoma & 470 & 64 \\
UCEC & Endometrial Carcinoma of the Uterine Corpua & 545 & 192 \\
\hline
\multicolumn{2}{l}{Total} & 5,060 & 576 \\
    \hline
  \end{tabular}
  \caption{The main contribution of our work: nucleus segmentation data in 10 cancer types. We also generated results in 4 additional cancer types (COAD: colon adenocarcinoma, READ: rectal adenocarcinoma, STAD: stomach adenocarcinoma, UVM: Uveal Melanoma) that are not as good as the 10 cancer types. To validate the segmentation data, we collected  segmentation ground truth in \textbf{1,356} patches. This set of manually segmented data is another contribution of our work.}
  \label{tab:data_results}
\end{table}

\section*{Methods}
\label{sec:methods}
\revision{We first describe our published nucleus segmentation method in the first subsection ``robust nucleus segmentation'', then describe the new quality control and data validation approaches for this work through the rest of this paper.}

\subsection*{Robust nucleus segmentation}
\label{sec:robust_nucleus_seg}
\begin{figure*}[th!]
\begin{center}
   \includegraphics[width=1.0\linewidth]{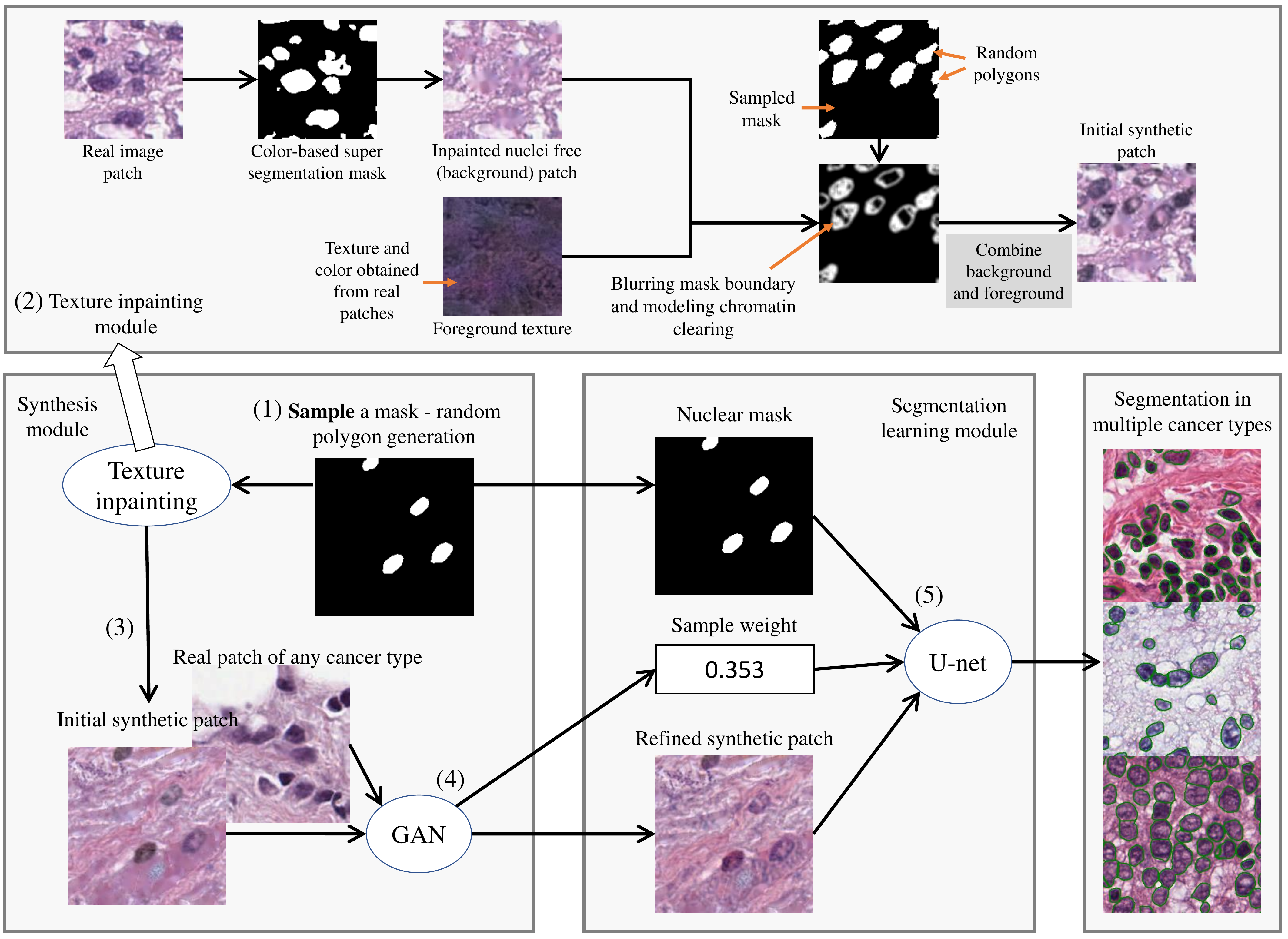}
\end{center}
   \caption{Overview of our nucleus segmentation model training: we use a texture inpainting module to synthesize an initial synthetic pathology image patch with its nuclear mask. We then refine the initial synthetic patch using a GAN and compute its sample weight. We finally train a segmentation CNN on this sampled instance. Details are in our technical paper \cite{hou2019robust} and source code repository.}
\label{fig:contribution}
\end{figure*}

To generate accurate segmentation results in multiple cancer types, existing state-of-the-art segmentation methods require extensive manually annotated training data in each cancer type. This is not scalable in practice. To address this problem, \revision{we use our existing robust nucleus segmentation model which was trained using not only manually annotated training data in several cancer types, but also heterogeneous synthetic training image patches, of every tissue type available in The Cancer Genome Atlas (TCGA).} This data synthesis method is unsupervised, and is capable of generating millions of training patches which normally requires thousands of human hours to manually annotate -- in this work, we used the data synthesis method to generate half a million patches. The workflow of this approach is shown in Figure \ref{fig:contribution}. We briefly describe our approach in this section.

We first generate possibly realistic nuclear masks as random polygons. Then, we construct an initial synthetic patch utilizing textures and colors from real tissue (texture inpainting module in Figure \ref{fig:contribution}). We then refine the initial synthetic patch, to make it more realistic. Along this process, we compute a sample weight of this synthetic patch, indicating how realistic it is. Finally, we train a segmentation network using the initially generated nuclear masks, refined synthetic patch, and sample weight. In other words, we enumerate possible ground truth structures first and then check if a resulting synthetic patch is realistic or not. We decrease its impact in the training loss if it is not realistic. Similarly, if a resulting patch is not only very realistic, but also rarely synthesized, then we increase its impact in the training loss. Details are described in our technical paper \cite{hou2019robust}.

In terms of the network architecture, the GAN's refiner has 21 convolutional layers and 2 pooling layers. The GAN's refiner discriminator has 15 convolutional layers and 3 pooling layers. As the segmentation CNNs, we use a U-net with 8 blocks: 4 down-sampling blocks and 4 up-sampling blocks. Each block has 3 to 6 convolutional layers and 1 pooling/deconv layer. We add a skip connection between blocks of the same resolution. In total there are 43 convolutional layers (including deconv). Each convolutional layer in the first and last block have 16 filters. After each pooling layer, we double the number of filters. We train the U-net on three real training datasets \cite{kumar2017dataset,vu2019methods} and our large scale synthetic dataset of 500,000 patches. The U-net has two output heads: one for nuclear center detection and one for nuclear material segmentation. We then apply the watershed method \cite{beucher1994watershed,bai2017deep} on detected centers and segmentation results, to output instance-level segmentation. During test time, we normalize stains \cite{reinhard2001color} in histopathology images before applying the U-net. We released our code on \href{http://github.com/SBU-BMI/quip_cnn_segmentation}{github}.

\subsubsection*{Comparing to other state-of-the-art segmentation methods}
\label{sec:comparing_unet}
Comparisons between our approach and other state-of-the-art level methods are detailed in our technical paper \cite{hou2019robust}. As a summary, on the MICCAI17 to MICCAI18 \cite{vu2019methods}, and Kumar \etal \cite{kumar2017dataset} datasets, U-net trained with synthetic and real training data achieved state-of-the-art level results, even though other comparable baseline methods \cite{chen2017dcan,hou2019sparse} use computationally more expensive models. \revision{For example, Mask R-CNN is 10 times more expensive compared to our U-net. In other words, we improve the performance of our segmentation method by adding synthetic training data, instead of increasing the neural network's capacity, which  would make the task of segmenting 5,060 WSIs computationally very expensive.}

\subsection*{Quality control and data validation approaches overview}
\begin{figure*}[th!]
\begin{center}
   \includegraphics[width=1.0\linewidth]{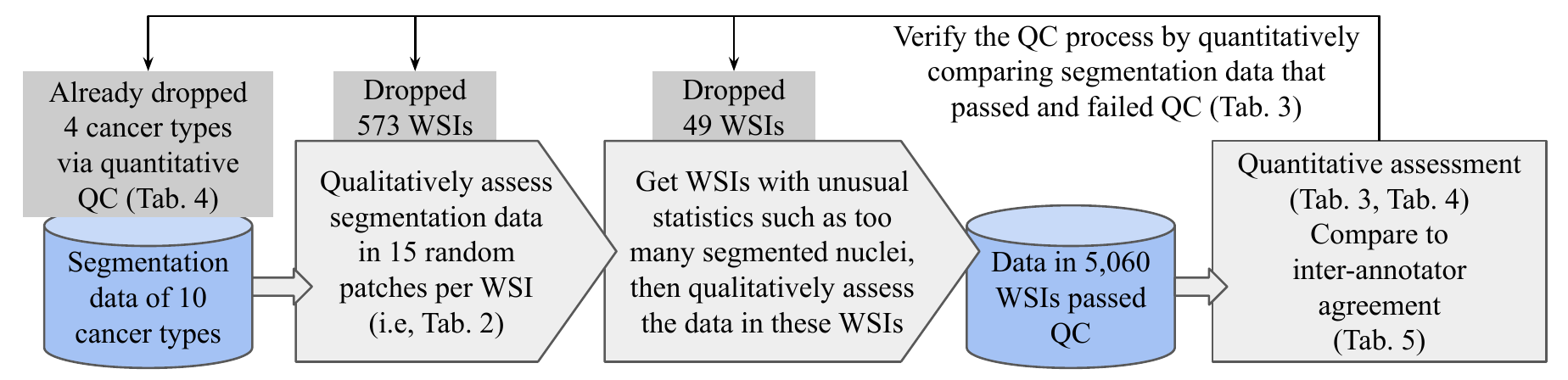}
\end{center}
   \caption{\revision{Our quality control and data validation pipeline. This QC process is implemented to evaluate segmentation results at the WSI level. It would be infeasible to check the segmentation quality of  all the  nuclei individually.}}
\label{fig:qc_pipeline}
\end{figure*}

\revision{We apply a Quality Control (QC) and evaluation process as shown in Figure \ref{fig:qc_pipeline}. This QC process is implemented to evaluate segmentation results at the WSI level, as it would be infeasible to perform quality-control on all nuclei individually. We focus our efforts on whole slide images from 10 tumor types after our initial qualitative QC led us to eliminate four cancer types. After the application of the QC process, there are 5,060 WSIs with acceptable segmentation results. The number of segmented nuclei in these WSIs is roughly 5 billion in total.}

\subsection*{WSI-level quality control}
We visually assess segmentation quality per WSI, and categorize WSIs into groups with different segmentation quality levels. It is very time consuming to go through each WSI: visually checking segmentation results in one WSI takes approximately 2.5 minutes; and thus 5,000 WSIs would require over 200 hours. Therefore, we sample segmentation data in each WSI-level in two ways:

\subsubsection*{Random segmentation region checking for quality control and rating}
We check segmentation quality in regions of all 5,060 WSIs at random locations. First, we randomly sample 15 patches \revision{(each has 256 by 256 pixels in 40X)} per WSI and mix all patches from all WSIs. This results in approximately \revision{76,000} patches. Then, we go through those patches and mark patches with reasonable segmentation results (both precision and recall are at least 75\%). Finally, we categorize WSIs into four groups, according to the number of patches with bad segmentations, as shown in Table \ref{tab:wsi-groups-define}.

\subsubsection*{WSI-level qualitative assessment}
\revision{The goal of this assessment is to identify and eliminate WSIs with unacceptable results. While this QC step involves a subjective method (i.e., visual inspection), it provides a complementary mechanism to the other QC steps (see Figure \ref{fig:qc_pipeline}). Unacceptable segmentation data identified in this way are still made available for download, but marked as ``failed WSI-level visual QC''.}

To make sure that we identify most slides with unacceptable segmentation results, we select slides that have unusual segmentation statistics for visual assessment. We visually assess segmentation results in these slides and mark slides with unacceptable results efficiently for quality control. We define ``unusual segmentation statistics'' as the following:
\begin{enumerate}
  \item Too many/few segmented nuclei. \revision{WSIs with either too many or too few segmented nuclei are subject to this WSI-level visual QC.}
  \item Average size of segmented nuclei is too large/small. \revision{WSIs with either very small or very large segmented nuclei are subject to this WSI-level visual QC.}
  \item Variation of the size of segmented nuclei is too large. \revision{WSIs with either very low or high nuclear pleomorphism are subject to this WSI-level QC.}
\end{enumerate}
In particular, we first compute the predicted nuclei count and average/variation of nuclear size, for each segmented slide. Then, slides that have one or more statistical values larger/smaller than $2\%$ of the slides within the same cancer type are selected for visual assessment using the caMicroscope web tool \cite{saltz2017containerized}. For a WSI, we rate the segmentation result in the slide as either acceptable or unacceptable. Following the random segmentation region checking criterion, it is acceptable if and only if in at least 80\% of the slide both precision and recall are at least 75\%. \revision{We check whether the segmentation data is above the threshold by visual assessment.} Around 500 WSIs in total are selected for visual assessment. For each cancer type, if a significant portion of the selected slides has unacceptable results, we select another $2\%$ (in total 4\%) of slides in each statistic value for visual assessment. In this way, \textbf{49} more slides were marked having unacceptable segmentations. Slides with results marked as unacceptable are excluded from analysis in the rest of this work.

We categorize WSIs into different levels of segmentation quality using random segmentation region checking and WSI-level visual assessment results, as summarized in Table \ref{tab:wsi-groups-define}.
\begin{table}
  \centering
  \begin{tabular}{rrr}
    \hline
               & Percentage of patches &  \\
    WSI groups & with bad segmentations & \#. slides \\
    \hline
    Best & 0\% & 2,346 \\
    Good & 0.01 - 6.67\% & 1,246 \\
    Adequate & 6.68 - 13.3\% & 593 \\
    Problematic & 13.4 - 20.0\% & 302 \\
    \multirow{2}{*}{Unacceptable} & $>$ 20.0\% & \multirow{2}{*}{573} \\
     & or failed WSI QC &  \\
    \hline
  \end{tabular}
  \caption{We categorize WSIs into groups with different segmentation quality levels. Slides identified as having unacceptable segmentation results are excluded from analysis in the rest of this work.}
  \label{tab:wsi-groups-define}
\end{table}

\subsection*{Patch-level manual annotation data}
\label{sec:patch_eval}
To quantitatively evaluate and validate the automatic segmentation results in each WSI group, we collect segmentation ground truth in 1,356 patches, uniformly distributed in 14 cancer types. Examples of manual segmentations are shown in Figure \ref{fig:wsi-seg-results}. \revision{All patches are 256 $\times$ 256 pixels in 40X (0.25 microns per pixel).} Since this dataset is large and contains 14 cancer types, we argue that it is a contribution of our work as well. To collect this large scale ground truth data, three graduate students, supervised by two pathologists, manually corrected automatic segmentation results given by a Mask R-CNN (detailed later in this section). Our manual segmentation is imperfect. However, its accuracy is only rarely limited by atypical chromatin patterns or representation of the entire nucleus in the plane of section, and rarely encompasses more than a portion of the nuclear contour. The imperfection level of manual segmentation results fell roughly within the range of variability that one would expect when one compares data from different human annotators - the Dice scores of both cases are within the range of 0.75 to 0.80.

Using this patch level segmentation ground truth, we evaluate the quality of our automatic segmentation data in each cancer type. We found that our results in 10 out of the 14 cancer types are relatively accurate. We release our segmentation data in those 10 cancer types as our main contribution (Table \ref{tab:data_results}).

\subsubsection*{Ground truth collection}
We first extract patches of $256 \time 256$ pixels in 40X, randomly (unbiased) and uniformly distributed in 14 cancer types. We label extracted patches in two ways, described below.

\paragraph{Fast manual segmentation by correcting Mask R-CNN's segmentation results.} In order to label thousands of patches, we minimize human labor by utilizing a Mask R-CNN - human annotators manually correct the Mask R-CNN's segmentation results in each patch, instead of labeling from scratch. Mask R-CNN \cite{he2017mask} is a state-of-the-art level instance level segmentation network which although is not computationally efficient for segmenting over thousands of slides, gives reasonable segmentation results. Another advantage of using Mask R-CNN is that it has a different architecture compared to the U-net that we use to generate segmentation results. This architectural different eliminates possible biases for evaluation. In particular, we use the authors \href{https://github.com/facebookresearch/maskrcnn-benchmark}{implementation} and train a Mask R-CNN on the same real + synthetic dataset used for training the U-net. We then apply the trained Mask R-CNN on 1,356 patches. Three graduate students then correct the segmentation results by 1). Segmenting unsegmented nuclei; 2). Removing false segmentations; 3). Modifying incorrect segmentations. Manual segmentation results are reviewed by two pathologists and patches significantly mislabeled are then relabeled. This process is a form of crowdsourcing \cite{amgad2019structured}.

\paragraph{Manual segmentation from scratch.} In order to evaluate the level of approximation in manual segmentation and the methodology of correcting Mask R-CNN's segmentation results, each of the three graduate students manually label a common set of 27 patches from scratch (not by correcting the Mask R-CNN's results). As a result, each patch has three manual segmentations, one from each student. Manual segmentation results are also reviewed by two pathologists and patches significantly mislabeled are then relabeled. Note that these patches were sampled from the same 1,356 patches described before.

\subsection*{Code availability}
Source code is available at on \href{github.com/SBU-BMI/quip_cnn_segmentation}{github}. It contains the following repositories:\\
\textbf{training-data-synthesis} Code for generating synthetic training data for nucleus segmentation model training.\\
\textbf{training-data-real-patch-extraction} Code for converting the format of real training data.\\
\textbf{segmentation-of-nuclei} Code for training a nucleus segmentation model on patches generated by the above-mentioned repositories, and applying a trained model on WSIs.

Detailed descriptions are in the README files in the Github repository. We also provide a Dockerfile in Github, containing a trained model for easy deployment.

\section*{Data Records}
All data records are included in The Cancer Imaging Archive (TCIA) \cite{hou2019dataset}.

\subsubsection*{Automatic nucleus segmentation data}
The algorithm-generated segmentation results. For each cancer type, you can find a cancertype\_polygon folder, for example, BLCA\_polygon. It contains polygon coordinates for each segmented nucleus (csv files), for all WSIs of BLCA. These results are obtained by thresholding the grayscale results in BLCA\_prob folder and separating touching or overlapping nuclei by combining the detection and segmentation results. Each line in a csv file contains information of one nucleus. There are three columns in a csv file:
\begin{description}
\item[$\bullet$~AreaInPixels] Size of the nucleus in terms of the number of pixels.
\item[$\bullet$~PhysicalSize] The number of pixels projected to 40X.
\item[$\bullet$~Polygon] The contour of the nucleus (polygon vertices in [x0:y0:x1:y1:..]).
\end{description}
In addition to cancertype\_polygon folders, there are cancertype\_meta folders which contain meta-data for each WSI. These folders are useless unless you use \href{https://github.com/camicroscope/caMicroscope}{caMicroscope} to visualize data.

Note: (1) In Box.com, the number of files under each folder shown in the ``size'' column is approximate; (2) Whether a slide has Unacceptable segmentation result or not is listed in the ``list of histopathology slides'' data described later. To further recognize WSIs with Best/good/Adequate/Problematic segmentations, one can use the ``random segmentation region checking result'' data described later.

\subsubsection*{List of histopathology slides}
The list of 5,060 WSIs and summarized quality control results. This is a csv file with the following columns:
\begin{description}
\item[$\bullet$~CancerType] Cancer type of the WSI.
\item[$\bullet$~WSI-ID] The case ID of the WSI, in TCGA naming convention.
\item[$\bullet$~QCResult] The summarized quality control result (passed or failed).
\end{description}

We do not redistribute the actual WSIs. These gigapixel histopathology slides can be downloaded from the publicly available The Cancer Genome Atlas (TCGA) repository \cite{TCGAdataset}. For example, to download Urothelial carcinoma of the bladder (BLCA) slides, a user can:
\begin{enumerate}
    \item Visit \url{portal.gdc.cancer.gov/projects/TCGA-BLCA}
    \item Click on the ``Files'' link in the ``Diagnostic Slide'' row.
    \item Click on the ``Add All Files to Cart'' bottom.
    \item Go to your cart, and download all cart items.
\end{enumerate}

\subsubsection*{WSI quality control result}
The list of slides selected for quality control by visual assessment and the detailed quality control result. This is a csv file with the following new columns (we do not list columns that are already explained before):
\begin{description}
\item[$\bullet$~NumNucleiSample] The number of segmented nuclei in this WSI.
\item[$\bullet$~SizeOfNuclei-Average] The average size of nuclei.
\item[$\bullet$~SizeOfNuclei-Stddev] The standard deviation of the size of nuclei.
\item[$\bullet$~Note] The reason of selecting this WSI for visual assessment.
\item[$\bullet$~SegmentationUnacceptableOrNot] 0: acceptable; ? or 1: unacceptable.
\item[$\bullet$~VisualAssessmentComment] Verbal comments on this WSI.
\end{description}

\subsubsection*{Random segmentation region checking result}
The detailed result of random segmentation region checking for each WSI. This is a csv file with the following new columns:
\begin{description}
\item[$\bullet$~NumOfUnacceptableSegRegions] The number of unacceptable regions.
\item[$\bullet$~NumOfSampledRegions] The total number of visually assessed regions.
\end{description}

\subsubsection*{Manual segmentation data}
The png images of manual segmentation data. Contains original H\&E stained histopathology image patches, and instance-level segmentation masks. Additional information is in the readme.txt file of this data.

\section*{Technical Validation}
We visually assess segmentation results in randomly sampled Whole Slide Images (WSIs) and also quantitatively analysis segmentation quality using patch-level segmentation labels.

\subsection*{WSI-level qualitative evaluation.}
Qualitative evaluation on all segmented WSIs is impractical. We randomly select 328 WSIs uniformly from 10 cancer types - at least 32 WSIs per cancer type to evaluate qualitatively. We use the same evaluation criterion used in the quality control process. Segmentation results in each slide is categorized as either acceptable or unacceptable. It is acceptable if and only if in at least 80\% of the slide both precision and recall are at least 75\%.

Out of the \textbf{328} randomly selected WSIs, \textbf{15} were marked as having unacceptable results. This concludes that our segmentation results on vast majority of WSIs are acceptable. We show examples of segmentation results in relatively large histopathology image tiles in Figure \ref{fig:wsi-seg-results}.

\subsection*{Patch-level quantitative evaluation}
We use manually annotated patches for quantitative evaluation. Note that we only use 971 patches in 10 cancer types, out of the 1,356 manually segmented patches in 14 cancer types. We only use manual segmentation in the center 226 $\times$ 226 pixels in each patch (as opposed to the entire 256 $\times$ 256 pixel patch), since segmentation close to the boundary is ambiguous due to incomplete data.

\subsubsection*{Evaluation metric}
\begin{figure}
\begin{center}
   \includegraphics[width=1.0\linewidth]{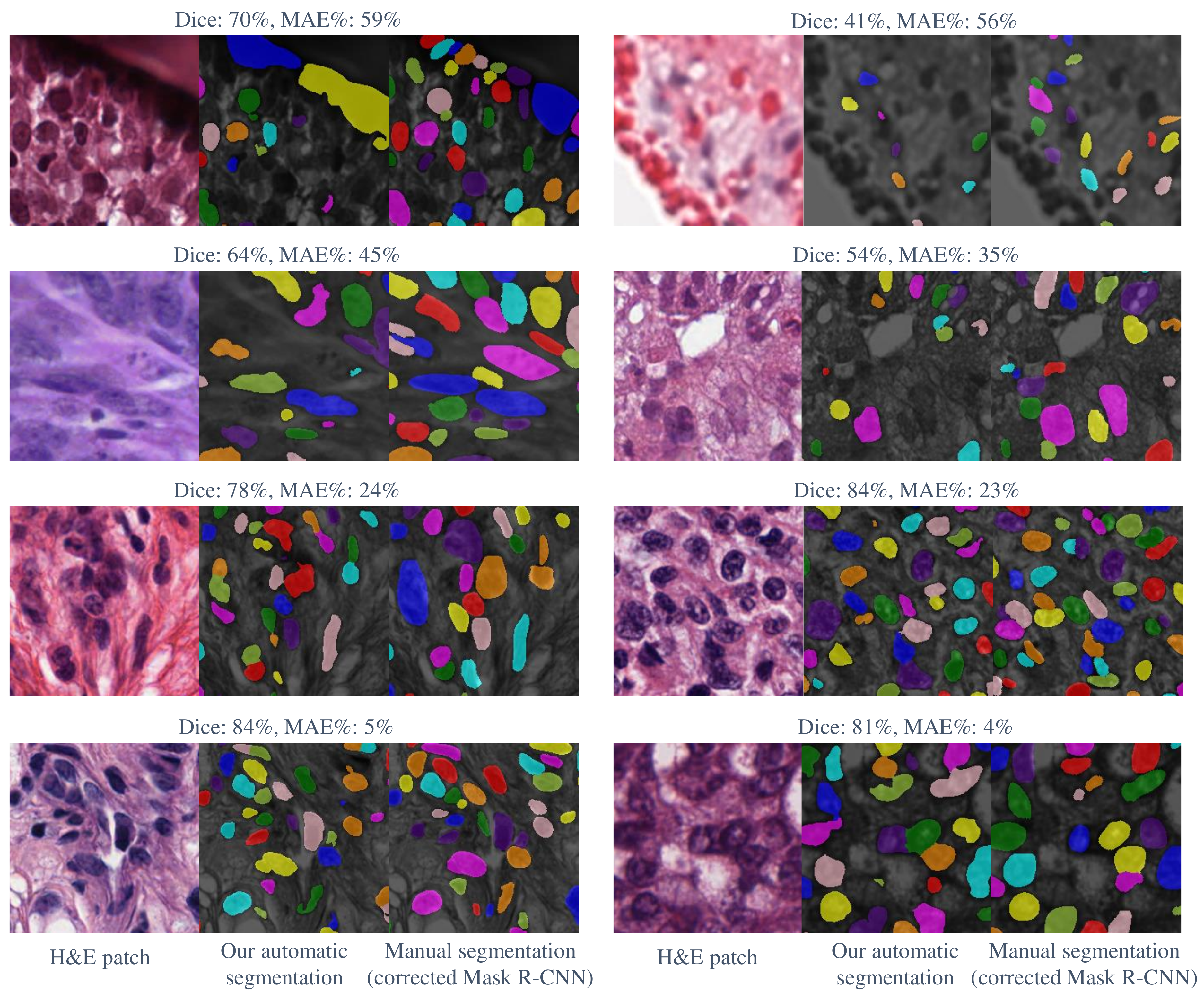}
\end{center}
   \caption{Examples of automatic segmentation vs. manual segmentation. First two rows: failure cases. Last two rows: randomly selected samples.}
\label{fig:good-bad-patch-pred}
\end{figure}

We use the Dice coefficient for measuring the quality of class-level (nuclear material or not) segmentation. Dice is ill-defined in patches that do not have any ground truth or predicted segmentation. To address this problem, the final Dice score is the average of per-patch Dice scores, weighted by the number of nuclei (ground truth nuclei count $+$ predicted nuclei count) in each patch. To jointly measure the quality of segmentation and the quality of separating individual nuclei, we use the Instance-Dice score which is also used in the MICCAI nucleus segmentation challenge \cite{vu2019methods}. In addition, we compute the Pearson correlation and Mean Absolute Error Ratio (MAE\%) between the number of nuclei segmented by U-net (defined as $p$), against the number of nuclei segmented by human annotators (defined as $t$). The MAE\% is computed below:
\begin{equation}
    \text{MAE\%} = \dfrac{\lvert p - t \rvert}{t} \text{,}
\end{equation} When we compute MAE\% on a set of patches, we first compute the average of $\lvert p - t \rvert$ and $t$ across all patches, then compute their ratio. We show examples of segmentation data with their evaluation results in Figure \ref{fig:good-bad-patch-pred}.

\begin{table}
  \centering
  \begin{tabular}{l r r r r r}
    \hline
            & \#. patch &           & Instance- & \multicolumn{2}{c}{Nuclei count} \\
 WSI groups & labels    & Dice      & Dice     & Correlat.         & MAE\% \\
    \hline
 Best & 446 & 0.797 & 0.687 & 0.947 & 15.2\% \\
 Good & 242 & 0.789 & 0.660 & 0.930 & 16.1\% \\
 Adequate & 128 & 0.774 & 0.636 & 0.915 & 17.6\% \\
 Problematic & 52 & 0.788 & 0.625 & 0.879 & 20.5\% \\
 Unacceptable & 103 & 0.690 & 0.545 & 0.718 & 33.8\% \\
    \hline
 Excluding unacceptable & 868 & 0.790 & 0.667 & 0.932 & 16.2\% \\
    \hline
  \end{tabular}
  \caption{Quantitative assessment of the quality of nucleus segmentation, across 10 cancer types. The definition of WSI groups are given in Table \ref{tab:wsi-groups-define}. We exclude unacceptable segmentation results from analysis work in the rest of this paper.}
  \label{tab:unet-vs-corrected-maskrcnn}
\end{table}

\subsubsection*{Generated segmentation results vs. corrected Mask R-CNN's results}
\begin{figure}
\begin{center}
   \includegraphics[trim=100 50 100 40, clip, width=0.8\linewidth]{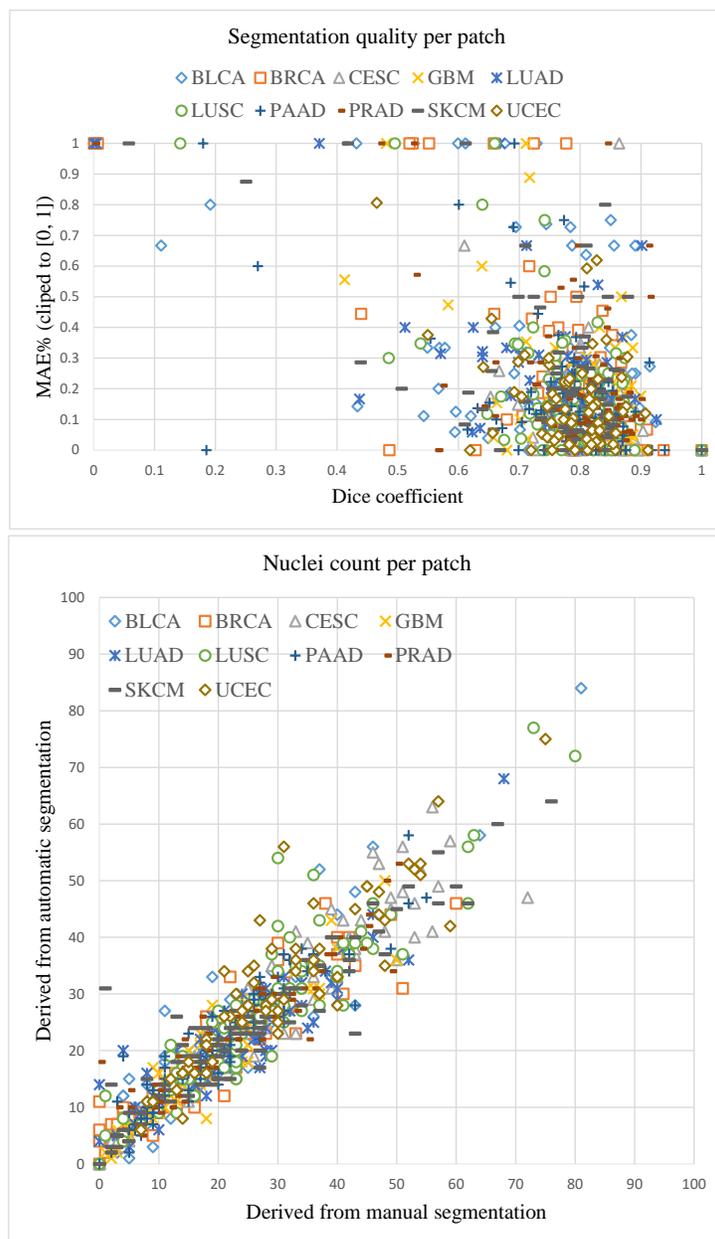}
\end{center}
   \caption{\textbf{Top:} Dice and MAE\% results of all patches. \textbf{Bottom:} Predicted nuclei count (derived from automatic segmentation) vs. Ground truth nuclei count (derived from manual segmentation). Pearson correlation $=0.932$, p-value $< 1.0\times 10^{-308}$.}
\label{fig:nuclei_count_scatter}
\end{figure}

We compare the automatic segmentation results with the manual segmentations obtained from correcting Mask R-CNN's results. The overall accuracy of generated segmentation results is shown in Table \ref{tab:unet-vs-corrected-maskrcnn}. A scatter chart (Figure \ref{fig:nuclei_count_scatter}) shows the accuracy of the predicted nuclei count. We also show per-cancer type evaluation results in Table \ref{tab:unet-vs-corrected-maskrcnn-per-cancer}.

\begin{table}
  \centering
  \begin{tabular}{l r r r r r}
    \hline
 Cancer & \#. patch &      & Instance- & \multicolumn{2}{c}{Nuclei count} \\
 Type   &    labels & Dice & Dice     & Correlat.   & MAE\%              \\
    \hline
BLCA & 95 & 0.779 & 0.668 & 0.941 & 20.5\% \\
BRCA & 89 & 0.798 & 0.649 & 0.922 & 19.6\% \\
CESC & 79 & 0.818 & 0.677 & 0.947 & 13.4\% \\
GBM & 86 & 0.809 & 0.723 & 0.938 & 14.4\% \\
LUAD & 88 & 0.772 & 0.641 & 0.896 & 17.4\% \\
LUSC & 97 & 0.789 & 0.665 & 0.924 & 16.1\% \\
PAAD & 91 & 0.785 & 0.679 & 0.933 & 15.8\% \\
PRAD & 96 & 0.799 & 0.670 & 0.940 & 14.7\% \\
SKCM & 86 & 0.774 & 0.675 & 0.933 & 17.1\% \\
UCEC & 61 & 0.778 & 0.629 & 0.900 & 14.6\% \\
    \hline
  \end{tabular}
  \caption{Quantitative assessment of the quality of nucleus segmentation, in each of the 10 cancer types. The p-value of Pearson correlation for every cancer type is smaller than $< 7.0\times 10^{-23}$.}
  \label{tab:unet-vs-corrected-maskrcnn-per-cancer}
\end{table}

\begin{table}
  \centering
  \begin{tabular}{p{3.7cm} r r r r}
    \hline
                 &      & Instance- & \multicolumn{2}{c}{Nuclei count} \\
 Inter-annotator & Dice & Dice      & Correlat.  & MAE\% \\
    \hline
 Annotator A vs. B & 0.760 & 0.600 & 0.959 & 10.8\% \\
 Annotator B vs. C & 0.752 & 0.622 & 0.959 & 15.5\% \\
 Annotator C vs. A & 0.774 & 0.697 & 0.954 & 12.2\% \\
    \hline
  \end{tabular}
  \caption{Agreements between annotations from different human annotators. This is the performance upper bond of any automatic segmentation method.}
  \label{tab:inter-human-seg-agreement}
\end{table}

\subsubsection*{Evaluating level of approximation in manual segmentation}
We evaluate the level of approximation in manual segmentation by comparing each annotator's segmentation result with each other. We apply the evaluation metrics between each pair of students, shown in Table \ref{tab:inter-human-seg-agreement}. One observation that in many cases, it is uncertain whether an object in histopathology images is a nucleus or not. This also contributes to the segmentation disagreement between human annotators.

\subsubsection*{Labeling from scratch vs. correcting Mask R-CNN's results}
Finally, we evaluate how the labeling from scratch vs. correcting Mask R-CNN's results differ. For the 27 patches that were labeled from scratch, there are also the Mask R-CNN's corrected results. Evaluation results are in Table \ref{tab:seg-scratch-vs-seg-maskrcnn}.

\begin{table}
  \centering
  \begin{tabular}{l r r r r}
    \hline
           &      & Instance- & \multicolumn{2}{c}{Nuclei count} \\
 Annotator & Dice & Dice      & Correlat.  & MAE\%  \\
    \hline
 Annotator A & 0.803 & 0.664 & 0.962 & 12.4\% \\
 Annotator B & 0.793 & 0.631 & 0.984 & 11.2\% \\
 Annotator C & 0.780 & 0.683 & 0.973 & 9.5\% \\
    \hline
  \end{tabular}
  \caption{Comparing labeling from scratch vs. correcting Mask R-CNN's results.}
  \label{tab:seg-scratch-vs-seg-maskrcnn}
\end{table}

\section*{Usage Notes}
We use CC0 (no copyright reserved) for our data.

Due to implementation and memory limitations, automatic nucleus segmentation results were generated and stored in 4,000 by 4,000 pixel tiles, as supposed to the entire WSI. Thus, nuclei across multiple tiles are split into different tiles. Additionally, we do not segment nuclei in tiles whose width or height is less than 2,000 pixels (this might happen on the edge of a WSI). All validation results include these by-design errors.

\section*{Acknowledgements}
This work was supported in part by 1U24CA180924-01A1, 3U24CA215109-02, and 1UG3CA225021-01 from the National Cancer Institute, R01LM011119-01 and R01LM009239 from the U.S. National Library of Medicine. This work leveraged resources from XSEDE, which is supported by NSF ACI-1548562 grant, including the Bridges system (NSF ACI-1445606) at the Pittsburgh 
Supercomputing Center.

\section*{Author contributions}
Conceptualization, J.H.S., L.H., T.M.K., D.S.; Methodology, L.H., J.H.S., D.S., T.M.K.; Investigation, L.H., J.H.S., R.G., T.M.K., Y.Z., K.S.; Writing, L.H., J.H.S., R.G., T.M.K.; Supervision, J.H.S., R.G., D.S., T.M.K., L.H.; Visualization, L.H., J.H.S; Data Curation, L.H., R.G., K.S., Y.Z.; Software, L.H., T.K.; Formal Analysis; L.H., J.H.S., R.G.

\section*{Competing interests}
The authors declare no competing interests.


\end{document}